\documentclass[12pt]{article}

\input epsf
\usepackage[dvips]{graphicx}
\usepackage{bm,latexsym,amsmath,amssymb,amsfonts,mathrsfs}
\usepackage{color}

\begin{document}

\begin{flushright}
RESCEU-40/13
\end{flushright}

\centerline{\Large \bf Derivative interactions}
\vskip 0.4cm
\centerline{\Large \bf in de Rham-Gabadadze-Tolley massive gravity}
\vskip 0.2cm
\centerline{\Large \bf }

\vskip 0.7cm
\centerline{\large Rampei Kimura and Daisuke Yamauchi}
\vskip 0.3cm

\centerline{\em Research Center for the Early Universe,}
\centerline{\em The University of Tokyo, Tokyo, 113-0033, Japan}

\vskip 1.9cm

\begin{abstract}
We investigate the possibility of a new massive gravity theory
with derivative interactions as an extension of 
de Rham-Gabadadze-Tolley massive gravity.
We find the most general Lagrangian of 
derivative interactions 
using Riemann tensor
whose cutoff energy scale 
is $\Lambda_3$, which is consistent with 
de Rham-Gabadadze-Tolley massive gravity.
Surprisingly, this infinite number of derivative interactions
can be resummed with the same method in
de Rham-Gabadadze-Tolley massive gravity,
and remaining interactions contain only two parameters.
We show that the equations of motion for 
scalar and tensor modes in the decoupling limit 
contain fourth derivatives with respect to spacetime,
which implies the appearance of ghosts at $\Lambda_3$.
We claim that consistent derivative interactions in
de Rham-Gabadadze-Tolley massive gravity have a mass scale $M$,
which is much smaller than the Planck mass $M_{\rm Pl}$.
\end{abstract}

\newpage

\section{Introduction}
The construction of consistent theories of a massive spin-2 field
has attracted a considerable attention 
from theoretical physicists 
since Fierz and Pauli proposed the linearized 
massive gravity in 1939 \cite{FP}. 
This theory consists of the linearized Einstein-Hilbert term 
and the quadratic mass term described by the fluctuation tensor $h_{\mu\nu}$
around Minkowski spacetime,
$g_{\mu\nu}=\eta_{\mu\nu}+h_{\mu\nu}/M_{\rm Pl}$.
The action of Fierz-Pauli massive gravity
is given by
\begin{eqnarray}
  S_{\rm FP}=\int d^4x 
  \biggl[
  -{1 \over 4}h^{\mu\nu}{\cal E}^{\alpha\beta}_{\mu\nu}h_{\alpha\beta}
  -m^2 (h_{\mu\nu}h^{\mu\nu}-h^2)
  +{1 \over M_{\rm Pl}}h_{\mu\nu}T^{\mu\nu}
  \biggr],
  \label{FPaction}
\end{eqnarray}
where ${\cal E}^{\alpha\beta}_{\mu\nu}$ 
is the Lichnerowicz operator,
$m$ is a graviton mass, 
$h$ is the trace of the fluctuation tensor $h_{\mu\nu}$,
and $T^{\mu\nu}$ is the energy-momentum tensor.
This mass term is constructed so that 
an extra ghost degree of freedom does not appear and
the number of propagating degrees of freedom is 5.
The massless limit of Fierz-Pauli theory
does not reproduce the result of general relativity;
i.e., the propagator of Fierz-Pauli theory, in the massless limit
does not give the massless one due to the coupling between
matter and the scalar mode of the graviton.
This problem is called van Dam-Veltman-Zakharov discontinuity \cite{vDV,Z}.
The question of this discontinuity was solved by 
Vainshtein \cite{Vainshtein}, 
who pointed out that nonlinearities become important
as the graviton mass goes to zero
and they recover general relativity through the so-called
Vainshtein mechanism.
Nonetheless, in Fierz-Pauli theory
the appearance of an extra ghost degree of freedom,
called the Boulware-Deser (BD) ghost,
is inevitable \cite{BD},
and thus the construction of consistent ghost-free theories 
has been thought to be impossible 
due to the lack of a Hamiltonian constraint \cite{GhostinMG}.
However, de Rham and Gabadadze constructed a ghost-free
theory in the so-called decoupling limit 
by adding appropriate combinations of nonlinear potential terms \cite{dRG}.
Furthermore, de Rham, Gabadadze, and 
Tolley (dRGT) showed that the infinite number of potentials can be resummed by
using a new tensor, which has a square root structure \cite{dRGT}.
The absence of the BD ghost in the full theory has been 
proven by Hassan and Rosen, and 
they confirmed the existence of a Hamiltonian constraint,
as well as a secondary constraint, 
which ensures the appropriate number of 
degrees of freedom of the massive graviton \cite{HR}.

Recently, Ref.~\cite{DI}  pointed out 
a possibility to add a `pseudo-linear' derivative interaction term\footnote{
These terms, Eq.(\ref{pseudo}), are not clearly linear in $h$, 
but they are the leading terms in the expansion
of small fluctuations,
which satisfies the gauge symmetry.
Therefore, we call them `pseudo-linear` derivative interactions.}
without introducing a new additional degree of freedom
in massive spin-2 theories;
it is given by
\begin{eqnarray}
  {\cal L}_{2,3} \sim M_{\rm Pl}^2 \, \varepsilon^{\mu\nu\rho\sigma}\varepsilon^{\alpha\beta\gamma\delta}
  \partial_{\mu}\partial_{\alpha}\,h_{\nu\beta}\,h_{\rho\gamma}\,h_{\sigma\delta},
\label{pseudo}
\end{eqnarray}
in four dimension.\footnote{
In Ref.~\cite{DI2}, 
the authors already derived the same pseudo-linear derivative interaction term, 
Eq.(\ref{pseudo}), 
in a different way before Ref.~\cite{DI}
appeared.
}
Here $\varepsilon^{\mu\nu\rho\sigma}$ is the Levi-Civita symbol 
normalized so that $\varepsilon^{0123}=-1$.
The antisymmetric structure of Eq.~(\ref{pseudo})
prevents $h_{00}$ from appearing nonlinearly; 
thus, this term is definitely linear in $h_{00}$, 
and it becomes a Lagrange multiplier,
which produces a Hamiltonian constraint.
However, $h_{00}$ itself does not give a Hamiltonian constraint
dRGT massive gravity; hence, we are not certain that
nonlinear generalizations of the derivative interaction Eq.~(\ref{pseudo}) exist.

In this paper,
we extend dRGT massive gravity theory
by constructing `nonlinear' derivative interactions
and we investigate whether these derivative interactions
are consistent or not.
In Sec 2, we briefly review dRGT massive gravity and $\Lambda_3$ theory 
in the decoupling limit.
In Sec 3, we derive the most general Lagrangian of 
nonlinear derivative interactions
using the Riemann tensor.
In Sec 4, we investigate the consistency 
of the nonlinear derivative interactions constructed
in Sec 3.
Section 5 is devoted to conclusions.

Throughout the paper, we use units in which the speed
of light and the Planck constant are unity, 
$c=\hbar=1$, 
and $M_{\rm Pl}$ is the reduced Planck mass related 
to Newton's constant by $M_{\rm Pl}=1/\sqrt{8 \pi G}$. 
We follow the metric signature convention $(-,+,+,+)$.
Some contractions of rank-2 tensors are denoted by
${\cal K}^{\mu}_{~\mu}=[{\cal K}]$,~
${\cal K}^{\mu}_{~\nu}{\cal K}^{\nu}_{~\mu}=[{\cal K}^2]$,~
${\cal K}^{\mu}_{~\alpha}{\cal K}^{\alpha}_{~\beta}{\cal K}^{\beta}_{~\mu}=[{\cal K}^3]$, 
and so on.

\section{de Rham-Gabadadze-Tolley massive gravity}
The action for ghost-free massive gravity is given by \cite{dRG,dRGT} 
\begin{eqnarray}
  S_{\rm MG}={M_{\rm Pl}^2\over 2}\int d^4x \sqrt{-g}\left[R
    -{m^2 \over 4}\left( {\cal U}_2+ \alpha_3 {\cal U}_3 + \alpha_4 {\cal U}_4\right)
  \right]+S_m[g_{\mu\nu}, \psi],
\end{eqnarray}
where the potentials are given by
\begin{eqnarray}
  {\cal U}_2&=&
  2\varepsilon_{\mu\alpha\rho\sigma}\varepsilon^{\nu\beta\rho\sigma}
  {\cal K}^{\mu}_{~\nu}{\cal K}^{\alpha}_{~\beta}
  =4\left([{\cal K}^2]-[{\cal K}]^2\right),
  \nonumber\\
  {\cal U}_3&=&
  \varepsilon_{\mu\alpha\gamma\rho}\varepsilon^{\nu\beta\delta\rho}
  {\cal K}^{\mu}_{~\nu}{\cal K}^{\alpha}_{~\beta}{\cal K}^{\gamma}_{~\delta}
  =-[{\cal K}]^3+3[{\cal K}][{\cal K}^2]
  -2[{\cal K}^3],
  \nonumber\\
  {\cal U}_4&=&
  \varepsilon_{\mu\alpha\gamma\rho}\varepsilon^{\nu\beta\delta\sigma}
  {\cal K}^{\mu}_{~\nu}{\cal K}^{\alpha}_{~\beta}
  {\cal K}^{\gamma}_{~\delta}{\cal K}^{\rho}_{~\sigma}\\
  &=&-[{\cal K}]^4+6[{\cal K}]^2[{\cal K}^2]
  -3[{\cal K}^2]^2-8[{\cal K}][{\cal K}^3]+6[{\cal K}^4],\nonumber
\end{eqnarray}
and 
\begin{eqnarray}
  {\cal K}^{\mu}_{~\nu}
  &=&\delta^{\mu}_{~\nu} -\sqrt{\delta^{\mu}_{~\nu} -H^{\mu}_{~\nu} }\nonumber\\
  &=&\delta^{\mu}_{~\nu} -\sqrt{\eta_{ab}g^{\mu\alpha}\partial_{\alpha}\phi^a\partial_{\nu}\phi^b}.
\label{Ktensor}
\end{eqnarray}
Here 
$\alpha_3$ and $\alpha_4$ are constants,
$\varepsilon_{0123}=\sqrt{-g}$,
the fluctuation tensor $H_{\mu\nu}$ is defined by
$H_{\mu\nu}=g_{\mu\nu}-\eta_{ab}\partial_{\mu}\phi^a\partial_{\nu}\phi^b$,
and $\phi^a$ is called the St{\" u}ckelberg field, 
which is responsible for restoring general covariance of the theory \cite{AGS}. 
The choice of the St{\" u}ckelberg field is arbitrary, and 
fixing the unitary gauge, $\phi^a=\delta^a_\mu x^\mu$, 
it reduces to Fierz-Pauli massive gravity at linear level. 

The decoupling limit is very convenient to capture 
high energy behavior below the graviton Compton wavelength.
Due to the decoupling of vector modes, 
we can safely ignore the vector modes in the decoupling limit.
Usually the St{\" u}ckelberg field can be expanded around
the unitary gauge,
\begin{eqnarray}
  \phi^a=\delta^a_\mu x^\mu -\eta^{a\mu} {\partial_\mu \pi / M_{\rm Pl}m^2},
\label{stuckelbergDL}
\end{eqnarray}
where $\pi$ describes the scalar mode of the massive graviton. 
We also expand the physical metric as $g_{\mu\nu}=\eta_{\mu\nu}+h_{\mu\nu}/M_{\rm Pl}$.
Thus, we can extract the tensor and scalar modes of the massive graviton
by taking the following limits,
\begin{eqnarray}
  M_{\rm Pl} \to \infty, \qquad m \to 0, 
  \qquad \Lambda_3 = (M_{\rm Pl}m^2)^{1/3}={\rm fixed}, 
  \qquad {T_{\mu\nu} \over M_{\rm Pl}}={\rm fixed}.
\end{eqnarray}
Then the action in the decoupling limit is given by
\begin{eqnarray}
  &&{\cal L}_{\rm DL}=
  -{1\over 4} h^{\mu\nu}{\cal E}_{\mu\nu}^{\alpha\beta}h_{\alpha\beta}
  -h^{\mu\nu} \biggl[{1\over 4} 
    \varepsilon_\mu^{~\rho\gamma\alpha}\varepsilon_{\nu\rho\gamma}^{~~~\beta} 
    \Pi_{\alpha\beta}
    +{3\alpha_3+4 \over 16 \Lambda_3^3}
    \varepsilon_\mu ^{~\gamma\alpha\rho}\varepsilon_{\nu \gamma}^{~~\beta\sigma}
    \Pi_{\alpha\beta}\Pi_{\rho\sigma}
\nonumber\\
  &&~~~~~~~~~~~~~~~~~~~~~~~~~~~~~~~~~
    +{\alpha_3+4\alpha_4\over 16\Lambda_3^6}
    \varepsilon_\mu ^{~\alpha\gamma\rho}\varepsilon_{\nu}^{~\beta\delta\sigma}
    \Pi_{\alpha\beta}\Pi_{\gamma\delta}\Pi_{\rho\sigma}\biggr]
+{1 \over M_{\rm Pl}}h^{\mu\nu}T_{\mu\nu},
\label{dRGTDLLagrangian}
\end{eqnarray}
where we defined $\Pi_{\mu\nu}\equiv \partial_\mu\partial_\nu\pi$. 
The $\Lambda_3$ is the cutoff energy scale of this theory,
and the theory above $\Lambda_3$ cannot be trusted.
The self-interactions of the scalar mode become the total derivative in 
the decoupling limit; therefore, the BD ghost does not appear at nonlinear level.
In addition, it is obvious that the remaining equations of motion 
for both $h_{\mu\nu}$ and $\pi$
are the second order differential equations,
which prevent the BD ghost from appearing in the theory.

\section{Construction of Lagrangians}
Now we want to construct nonlinear derivative interactions 
in dRGT massive gravity.
To this end, we demand the following restrictions :

\begin{enumerate}
 \item Linearlization of $h_{\mu\nu}$ reproduces Fierz-Pauli massive gravity. 
 \item The cutoff energy scale is $\Lambda_3$. 
 \item A derivative interaction term should contribute 
   at the energy scale $\Lambda_3$.
 \item The resultant theory does not have a BD ghost.
\end{enumerate}
There are a number of candidates for nonlinear derivative interactions,
such as\footnote{Here we restrict the form of derivative interactions
by using only a Riemann tensor. 
There might be other derivative interactions using the covariant derivative of 
$H_{\mu\nu}$.}
\begin{eqnarray}
  {\cal L}_{\rm int} \supset { M_{\rm Pl}^2 \sqrt{-g} H R,\, M_{\rm Pl}^2 \sqrt{-g} H^2 R,
    \,  M_{\rm Pl}^2\sqrt{-g} H^3 R, \,\cdot\cdot\cdot}.
\end{eqnarray}
Here we set the mass scale to be $M_{\rm Pl}^2$ for requirement 2, 
as we will see later.
First, we count the energy scales in the decoupling limit.
From Eq.~(\ref{stuckelbergDL}),
$H_{\mu\nu}$ undergoes the following transformation,
\begin{eqnarray}
  H_{\mu\nu} \to {h_{\mu\nu} \over M_{\rm Pl}}
  +2{\partial_{\mu}\partial_{\nu}\pi \over M_{\rm Pl}m^2}
  -{\partial_{\mu}\partial_{\alpha}\pi \partial_{\nu}\partial^{\alpha}\pi
    \over M_{\rm Pl}^2m^4},
\end{eqnarray}
and then the canonically normalized Lagrangian can be schematically written as
\begin{eqnarray}
  {\cal L}_{\rm int} \sim \Lambda_\lambda^{2-n_h-3n_\pi} h^{n_h-1} \partial^2 h\, (\partial^2\pi)^{n_\pi},
\end{eqnarray}
where we defined the energy scale 
\begin{eqnarray}
  \Lambda_\lambda=(M_{\rm Pl}m^{\lambda-1})^{1/\lambda}, \qquad 
  \lambda = \frac{n_h+3n_\pi-2}{n_h+n_\pi-2}.
\label{energyScale}
\end{eqnarray}
Here $n_h \geq 1$ and $n_\pi \geq 1$.
For the lowest order of $h_{\mu\nu}$, $n_h=1$, 
the energy scales are 
$\Lambda_5$ for $n_\pi=2$,
$\Lambda_4$ for $n_\pi=3$, 
and $\Lambda_{11/3}$ for $n_\pi=4$,
which are lower energy scales than $\Lambda_3$.
Therefore, in order to satisfy requirement 2, 
 $\partial^2 h\, (\partial^2\pi)^{n_\pi}$
has to be eliminated by the construction of 
the Lagrangian, and in the next section we show that 
such eliminations are possible for derivative interactions.
For the next order of $h_{\mu\nu}$, $n_h=2$,
the energy scale is always $\Lambda_3$ and 
does not depend on the value of $n_\pi$,
which automatically satisfies requirement 3.

\subsection{$HR$ order}
In this subsection we start with the lowest order terms in a general form,
\begin{eqnarray}
  {\cal L}_{{\rm int},1}=M_{\rm Pl}^2\sqrt{-g} H_{\mu\nu}(R^{\mu\nu}+ d\, R g^{\mu\nu}),
\end{eqnarray}
where $d$ is a constant. 
To determine the constant $d$, 
we first take the unitary gauge, $H_{\mu\nu}=h_{\mu\nu}/M_{\rm Pl}$, 
and linearize the Lagrangian around Minkowski spacetime,
$g_{\mu\nu}=\eta_{\mu\nu}+h_{\mu\nu}/M_{\rm Pl}$.
Then the lowest order of ${\cal L}_{{\rm int},1}$ gives the order of $(\partial h)^2$,
which is the same order of the quadratic Lagrangian of 
the Einstein-Hibert term in (\ref{FPaction}).
In order to satisfy requirement $1$,  
the quadratic action of ${\cal L}_{{\rm int},1}$ has to be
proportional to the Einstein-Hilbert term,
\begin{eqnarray}
  {\cal L}_{{\rm int},1}^{(2)}
  \propto 
\biggl[\sqrt{-g}R\biggr]_{h^2}.
\end{eqnarray}
Therefore, we require $d=-1/2$.
Then ${\cal L}_{{\rm int},1}$ can be written in terms of the Riemann dual tensor,
\begin{eqnarray}
  {\cal L}_{{\rm int},1}={1 \over 2}M_{\rm Pl}^2\sqrt{-g}\,
  \varepsilon^{\mu\nu\rho\sigma}\varepsilon^{\alpha\beta\gamma}_{~~~~\sigma}
  R_{\mu\alpha\nu\beta}\,H_{\rho\gamma}.
\end{eqnarray}

As we stated in the beginning of this section, 
the energy scale of $n_h=1$ terms in the decoupling limit is potentially 
dangerous and these terms have to be eliminated.
Therefore, we take the decoupling limit of the Lagrangian ${\cal L}_{{\rm int},1}$.
Using the property,
\begin{eqnarray}
\biggl[\sqrt{-g}
  \varepsilon^{\mu\nu\rho\sigma}\varepsilon^{\alpha\beta\gamma}_{~~~~\sigma}
  R_{\mu\alpha\nu\beta}\biggr]_{h}
=-{1\over M_{\rm Pl}}\varepsilon^{\mu\nu\rho\sigma}\varepsilon^{\alpha\beta\gamma}_{~~~~\sigma}
  \partial_{\mu}\partial_{\alpha}\,h_{\nu\beta},
\end{eqnarray}
the lowest order term for $n_h=1$ is given by
\begin{eqnarray}
  {\cal L}_{{\rm int},1}\biggr|_{\partial^2h \, \partial^2 \pi}
&=&-{1 \over m^2}\varepsilon^{\mu\nu\rho\sigma}\varepsilon^{\alpha\beta\gamma}_{~~~~\sigma}
  \partial_{\mu}\partial_{\alpha}\,h_{\nu\beta}\,\partial_{\rho}\partial_{\gamma}\pi\nonumber\\
&=&-{1 \over m^2}\partial_{\gamma}(\varepsilon^{\mu\nu\rho\sigma}\varepsilon^{\alpha\beta\gamma}_{~~~~\sigma}
  \partial_{\mu}\partial_{\alpha}\,h_{\nu\beta}\,\partial_{\rho}\pi).
\end{eqnarray}
This is nothing but a total derivative, and the elimination
of the $\partial^2 h\, \partial^2\pi$ order term 
is automatically satisfied by the antisymmetric structure of ${\cal L}_{{\rm int},1}$.
However, the next order $n_\pi=2$ is not a total derivative,
\begin{eqnarray}
  {\cal L}_{{\rm int},1}\biggr|_{\partial^2h \, (\partial^2 \pi)^2}
&=&{1 \over 2\Lambda_5^5}\varepsilon^{\mu\nu\rho\sigma}\varepsilon^{\alpha\beta\gamma}_{~~~~\sigma}
  \partial_{\mu}\partial_{\alpha}\,h_{\nu\beta}\,
  \partial_{\rho}\partial_{\kappa}\pi\partial^{\kappa}\partial_{\gamma}\pi.
\label{L1_d2}
\end{eqnarray}
The only way to eliminate this term is to add the next order Lagrangian,
\begin{eqnarray}
  {\cal L}_{{\rm int},1,2}={1 \over 8} M_{\rm Pl}^2\sqrt{-g}\,
  \varepsilon^{\mu\nu\rho\sigma}\varepsilon^{\alpha\beta\gamma}_{~~~~\sigma}
  R_{\mu\alpha\nu\beta}\,H_{\rho \kappa}H^\kappa_{~\gamma}.
\end{eqnarray}
This Lagrangian clearly produces the counterterm of Eq.(\ref{L1_d2}), but
it contains an $n_\pi=3$ term, 
\begin{eqnarray}
  {\cal L}_{{\rm int},1,2}\biggr|_{\partial^2h \, (\partial^2 \pi)^3}
&=&{1 \over 2\Lambda_4^8}\varepsilon^{\mu\nu\rho\sigma}\varepsilon^{\alpha\beta\gamma}_{~~~~\sigma}
  \partial_{\mu}\partial_{\alpha}\,h_{\nu\beta}\,
  \partial_{\rho}\partial_{\kappa}\pi\partial^{\kappa}\partial_{\lambda}\pi\partial^{\lambda}\partial_{\gamma}\pi.
\end{eqnarray}
This $n_\pi=3$ term can also be eliminated by adding the Lagrangian,
\begin{eqnarray}
  {\cal L}_{{\rm int},1,3}={1 \over 16} M_{\rm Pl}^2 \sqrt{-g}\,
  \varepsilon^{\mu\nu\rho\sigma}\varepsilon^{\alpha\beta\gamma}_{~~~~\sigma}
  R_{\mu\alpha\nu\beta}\,H_{\rho \kappa}H^\kappa_{~\lambda}H^\lambda_{~\gamma}.
\end{eqnarray}
Then we can use the same procedure 
to eliminate the  $n_h=1$ term in the decoupling limit
by introducing appropriate counterterms.
The Lagrangian is given by 
\begin{eqnarray}
  &&{\cal L}_{{\rm int},1}
= M_{\rm Pl}^2 \sqrt{-g}\,
  \varepsilon^{\mu\nu\rho\sigma}\varepsilon^{\alpha\beta\gamma}_{~~~~\sigma}
  R_{\mu\alpha\nu\beta}\nonumber \\
&&~~~~~~~~~~~
\times
  \biggl(
{1\over 2}H_{\rho \gamma}
+{1\over 8}H_{\rho \kappa}H^\kappa_{~\gamma}
+{1\over 16}H_{\rho \kappa}H^\kappa_{~\lambda}H^\lambda_{~\gamma}
+{5\over 128}H_{\rho \kappa}H^\kappa_{~\lambda}H^\lambda_{~\tau}H^\tau_{~\gamma}
+\cdot\cdot\cdot
\biggr)\nonumber \\
&&~~~~~~~~~~~
= - M_{\rm Pl}^2 \sqrt{-g}\,
  \varepsilon^{\mu\nu\rho\sigma}\varepsilon^{\alpha\beta\gamma}_{~~~~\sigma}
  R_{\mu\alpha\nu\beta}\,
g_{\rho\lambda}
\sum_{n=1}^{\infty}{\bar d}_n (H^n)^\lambda_{~\gamma},
\end{eqnarray}
where ${\bar d}_n$ is the expansion coefficient, 
and $(H^n)^\mu_{~\nu}\equiv H^\mu{}_{\alpha_1}H^{\alpha_1}{}_{\alpha_2}\cdot\cdot\cdot H^{\alpha_{n-1}}{}_{\nu}$.
One can notice that
the coefficients of these counterparts have the following recursive relation,
\begin{eqnarray}
  {\bar d}_n =- \sum_{i=1}^{i\leq n/2} (-1)^i\, 2^{-2i} \, {}_{n-i}{\rm C}_i \,{\bar d}_{n-i},
\end{eqnarray}
where the upper bound of the summation $n/2$ is chosen to be the largest integer,
${\bar d}_1=-1/2$ and ${}_n {\rm C}_r = n! / ((n-r)!r!)$.
This coefficient is nothing but the expansion coefficients of the ${\cal K}$ tensor,
${\bar d}_n=(2n)!/((1-2n)(n!)^2 4^n)$.
Using the expanded expression of (\ref{Ktensor}),
${\cal K}^{\mu}_{~\nu}=-\sum_{n=1}^{\infty}{\bar d}_n (H^n)^\mu_\nu$,
the Lagrangian can be resummed by using the ${\cal K}$  tensor,
\begin{eqnarray}
  {\cal L}_{{\rm int},1}=M_{\rm Pl}^2\sqrt{-g}\,
  \varepsilon^{\mu\nu\rho\sigma}\varepsilon^{\alpha\beta\gamma}_{~~~~\sigma}
  R_{\mu\alpha\nu\beta}\, {\cal K}_{\rho \gamma}.
\end{eqnarray}
This Lagrangian does not have the terms of the energy scales below $\Lambda_3$,
and nonlinear terms contribute at $\Lambda_3$.
Note that from the definition of the ${\cal K}$  tensor,
${\cal K}_{\mu\nu}|_{h_{\mu\nu}=0}\equiv \partial_\mu\partial_\nu\pi$,
we have only one $n_h=1$ term in the decoupling limit, and 
the ${\cal K}$  tensor ensures that the $n_h=1$ term 
is a total derivative in the decoupling limit.

\subsection{$H^2 R$ order}
Next we want to consider the next order Lagrangian, ${\cal O}(H^2 R)$.
Since we want to eliminate the energy scales below $\Lambda_3$,
we perform the same procedure as the  $HR$ case.
THe starting point of the Lagrangian is
\begin{eqnarray}
  {\cal L}_{{\rm int},2}={1\over 4}M_{\rm Pl}^2\sqrt{-g}\,
  \varepsilon^{\mu\nu\rho\sigma}\varepsilon^{\alpha\beta\gamma\delta}
  R_{\mu\alpha\nu\beta}\, H_{\rho \gamma} \, H_{\sigma \delta}.
\label{H2R}
\end{eqnarray}
This is the only total derivative combination which eliminates the 
$\partial^2h \, (\partial^2 \pi)^2$ term.
If we have different combinations of $H^2R$,
then the $\partial^2h \, (\partial^2 \pi)^2$ term remains
because higher order Lagrangians $H^3R$ cannot eliminate the $\partial^2h \, (\partial^2 \pi)^2$ term.
Now the term (\ref{H2R}) produces the $\partial^2h \, (\partial^2 \pi)^3$
term, but we can always add counterterms to eliminate
order by order. 
With the same procedure in the previous subsection, 
the counterterms can be resummed by using the ${\cal K}$ tensor again,
\begin{eqnarray}
  {\cal L}_{{\rm int},2}=M_{\rm Pl}^2\sqrt{-g}\,
  \varepsilon^{\mu\nu\rho\sigma}\varepsilon^{\alpha\beta\gamma\delta}
  R_{\mu\alpha\nu\beta}\, {\cal K}_{\rho \gamma} \, {\cal K}_{\sigma \delta}.
\label{RKK}
\end{eqnarray}
Apparently, the $\partial^2h \, (\partial^2 \pi)^2$ term are a total derivative,
and there is no higher order terms of $\pi$ for $n_h=1$ in the decoupling limit
from the definition of the ${\cal K}$ tensor.
Surprisingly, Linearization of (\ref{RKK}) in $h_{\mu\nu}$ in unitary gauge
gives the `pseudo-linear' derivative interaction term (\ref{pseudo}).

One might think that we can start with the ${\cal O}(H^3R)$ Lagrangian; 
however, we do not have a total derivative combination of 
$\partial^2h \, (\partial^2 \pi)^3$
due to the number of indices of the antisymmetric tensor in four dimensions,
which means there is no higher order Lagrangian
satisfying the restrictions.
Therefore, the most general Riemann derivative interaction 
for dRGT massive gravity is
\begin{eqnarray}
  {\cal L}_{\rm int}=\alpha M_{\rm Pl}^2\sqrt{-g}\,\varepsilon^{\mu\nu\rho\sigma}\varepsilon^{\alpha\beta\gamma}_{~~~~\sigma}
  R_{\mu\alpha\nu\beta}\, {\cal K}_{\rho \gamma}
+
\beta M_{\rm Pl}^2\sqrt{-g}\,
  \varepsilon^{\mu\nu\rho\sigma}\varepsilon^{\alpha\beta\gamma\delta}
  R_{\mu\alpha\nu\beta}\, {\cal K}_{\rho \gamma} \, {\cal K}_{\sigma \delta},
\label{DILagrangian}
\end{eqnarray}
where $\alpha$ and $\beta$ are model parameters.

\subsection{Arbitrary dimensions}
The extension to arbitrary dimensions can be 
easily done with the same method, and 
the Lagrangian is given by
\begin{eqnarray}
  {\cal L}_{\rm int}^{(D,d,n)}&=& 
  M_{\rm Pl}^{D-2}m^{2-d}\sqrt{-g} \,
  \varepsilon^{\mu_1\mu_2 \cdot\cdot\cdot \mu_D}\varepsilon^{\nu_1\nu_2\cdot\cdot\cdot\nu_D}
R_{\mu_{1}\nu_{1}\mu_{2}\nu_{2}} \cdot\cdot\cdot R_{\mu_{d-1}\nu_{d-1}\mu_d\nu_d} \, \\
  \nonumber
  &&\times \,   g_{\mu_{d+1}\nu_{d+1}} \cdot\cdot\cdot g_{\mu_n\nu_n}\, 
  {\cal K}_{\mu_{n+1}\nu_{n+1}} \cdot\cdot\cdot {\cal K}_{\mu_D\nu_D},
\label{DILagrangian}
\end{eqnarray}
where $D$ is the number of dimensions,
$d$ is an even number, and 
$n$ is an integer.
Here, $d/2$ is the number of the Riemann tensor,
$n-d$ is the number of the metric tensor,
and $D-n$ is the number of the ${\cal K}$ tensor,
satisfying  the relation,
$2 \leq d \leq m \leq D-1$.
Similarly, the lowest order of this term in the decoupling limit
always becomes a total derivative, so the contributions of the scalar mode 
in the decoupling limit are always at the energy scale $\Lambda_{(D+2)/(D-2)}\equiv (m^{4/(D-2)}M_{\rm Pl})^{(D-2)/(D+2)}$.
Note that in the unitary gauge description, 
these derivative interaction terms in $D$ dimensions
include the derivative interactions discussed in \cite{DI}.

\section{Equations of motion  of $\Lambda_3$ theory}
So far we could successfully eliminate all energy scales below $\Lambda_3$,
and the derivative interactions (\ref{DILagrangian}) 
in the decoupling limit contribute
at the energy scale $\Lambda_3$. 
Now we want to check whether these derivative interactions are 
free of a BD ghost or not in the decoupling limit.
For completeness, in the Appendix we show the lack of a Hamiltonian and 
momentum constraints in the Arnowitt-Deser-Misner formalism.

The Lagrangian ${\cal L}_{{\rm int},1}$ in the decoupling limit 
can be written as 
\begin{eqnarray}
  &&{\cal L}_{{\rm int},1}={1 \over \Lambda_3^3} \biggl[\sqrt{-g}\,\varepsilon^{\mu\nu\rho\sigma}\varepsilon^{\alpha\beta\gamma}_{~~~~\sigma}
  R_{\mu\alpha\nu\beta}\biggr]_{h\partial^2 h }
\, \partial_{\rho}\partial_{\gamma}\pi
\nonumber \\
&&~~~~~~~~~~~
-\sum_{n_\pi=1}^\infty {1 \over \Lambda_3^{3n_\pi}} \,
\varepsilon^{\mu\nu\rho\sigma}\varepsilon^{\alpha\beta\gamma}_{~~~~\sigma}
  \partial_{\mu}\partial_{\alpha}\,h_{\nu\beta} \,
\biggl[ {\cal K}_{\rho \gamma}\biggr]_{h(\partial^2 \pi)^{n_\pi}}.
\label{Lint1DL}
\end{eqnarray}
For $n_\pi=1$, the equation of motion for 
$h_{\alpha\beta}$ contains
third derivative terms, 
\begin{eqnarray}
  &&\frac{\delta{\cal L}_{{\rm int},1}}{\delta h^{\alpha\beta}}\biggl|_{h \partial^2h \, \partial^2 \pi}
  \supset
{1 \over \Lambda_3^3}
\biggl[
  {1 \over 4} (\partial_\alpha h_\beta^{~\mu} 
  + 2 \partial^\mu h_{\alpha\beta}-\partial_\beta h_\alpha^{~\mu})
  \square \partial_\mu \pi
  +{1 \over 2} (\partial^\nu h - \partial_\mu h^{\mu\nu})
  \partial_\nu \partial_{\alpha} \partial_\beta \pi \nonumber\\
  &&~~~~~~~~~~~~~~~~~~~~~~~~~~~~
  +{1 \over 4} (\partial_\beta h^{\mu\nu} - \partial^\nu h_{\beta}^{~\mu})
  \partial_\mu \partial_{\nu} \partial_\alpha \pi
  +{1 \over 4} (\partial_\alpha h^{\mu\nu} - \partial^\nu h_{\alpha}^{~\mu})
  \partial_\mu \partial_{\nu} \partial_\beta \pi\nonumber\\
  &&~~~~~~~~~~~~~~~~~~~~~~~~~~~~
  +{1 \over 2} \eta_{\alpha\beta}(\partial_\mu h^{\mu\nu} - \partial^\nu h) 
  \square \partial_\nu \pi
\biggr].
\end{eqnarray}
Here fourth derivative terms are accidentally canceled due to 
antisymmetric tensor. One can check that the equation of motion 
for $\pi$ also contains the third derivative of $h_{\mu\nu}$, 
and fourth derivative terms are canceled as well.
For $n_\pi=2$,  the equation of motion for $\pi$ 
contains the following term,
\begin{eqnarray}
  &&\frac{\delta{\cal L}_{{\rm int},1}}{\delta h^{\alpha\beta}}\biggl|_{h \partial^2h \, (\partial^2 \pi)^2}
\supset
-{1\over 4\Lambda_3^6}\,
\varepsilon^{\mu\rho\sigma}_{~~~~\alpha}\varepsilon^{\nu\gamma}_{~~~\sigma\beta}
\,h_{\rho \kappa} \,
\partial_{\mu}\partial_{\nu}\partial^{\kappa}\partial_{\lambda}\pi\,
\partial^{\lambda}\partial_{\gamma}\pi.
\end{eqnarray}
This equation of motion obviously has fourth derivative terms,
and one can easily show that the equation of motion of $h_{\mu\nu}$ 
for $n_\pi=2$
also contains fourth derivative terms. 
Therefore, a BD ghost appears at $\Lambda_3$
in ${\cal L}_{{\rm int},1}$.

The Lagrangian ${\cal L}_{{\rm int},2}$ in the decoupling limit 
can be written as 
\begin{eqnarray}
  &&{\cal L}_{{\rm int},2}={1 \over 4\Lambda_3^6} \biggl[\sqrt{-g}\,\varepsilon^{\mu\nu\rho\sigma}\varepsilon^{\alpha\beta\gamma\delta}
  R_{\mu\alpha\nu\beta}\biggr]_{h\partial^2 h }
\, \partial_{\rho}\partial_{\gamma}\pi\partial_{\sigma}\partial_{\delta}\pi
\nonumber\\
&&~~~~~~~~~~~
-\sum_{n_\pi=1}^\infty {1 \over \Lambda_3^{3n_\pi}} \,
\varepsilon^{\mu\nu\rho\sigma}\varepsilon^{\alpha\beta\gamma\delta}
  \partial_{\mu}\partial_{\alpha}\,h_{\nu\beta} \,
\biggl[ {\cal K}_{\rho \gamma} {\cal K}_{\sigma \delta}\biggr]_{h(\partial^2 \pi)^{n_\pi}}.
\end{eqnarray}
The lowest order term $n_\pi =1$ comes from the second term, and 
it is given by
\begin{eqnarray}
  {\cal L}_{{\rm int},2}\biggl|_{h \partial^2h \, \partial^2 \pi}&=&
-{1 \over \Lambda_3^3}
\,\varepsilon^{\mu\nu\rho\sigma}\varepsilon^{\alpha\beta\gamma\delta}
  \partial_{\mu}\partial_{\alpha}\,h_{\nu\beta} \,
h_{\rho\gamma}\partial_{\sigma}\partial_{\delta}\pi\nonumber \\
&=&
-{1 \over \Lambda_3^3}
\,\varepsilon^{\mu\nu\rho\sigma}\varepsilon^{\alpha\beta\gamma\delta}
  \partial_{\mu}\partial_{\alpha}\,h_{\nu\beta} \,
\partial_{\sigma}\partial_{\delta}h_{\rho\gamma}\,\pi\nonumber \\
&=& 
{1 \over \Lambda_3^3}\,
\pi \biggl[R^2-4R_{\mu\nu}R^{\mu\nu}+R_{\mu\nu\rho\sigma}R^{\mu\nu\rho\sigma}\biggr]_{h^2}.
\end{eqnarray}
This term is the Gauss-Bonnet term 
with nonminimal coupling of $\pi$
and does not yield higher order derivatives in the equation of motion.
Therefore a ghostly extra degree of freedom does not appear from this term.
This term is expected from a pseudo-linear derivative interaction 
in the decoupling limit as found in Ref.~\cite{DI}.
However, the equations of motion for $h_{\mu\nu}$ and $\pi$ 
of the next order $n_\pi =2$
contain third derivative terms, and the next order $n_\pi =3$
yields fourth derivative terms, similarly to $R{\cal K}$ term.
These higher derivative terms in the equations of motion 
cannot be eliminated by the choice of the parameters $\alpha$ and $\beta$.
Thus BD ghosts appear at $\Lambda_3$ 
in the derivative interactions in dRGT massive gravity theory.

To avoid BD ghosts from the derivative interactions (\ref{DILagrangian}),
the mass scale $M$ in the derivative interactions
has to be smaller than the Planck mass $M_{\rm Pl}$. 
Then the derivative interactions do not contribute at $\Lambda_3$ 
in the decoupling limit, and ghost modes appear
at the energy scale higher than the cutoff energy scale $\Lambda_3$.
In this case the Lagrangian in the decoupling limit
is  exactly given by the pure dRGT theory (\ref{dRGTDLLagrangian}).
It is interesting to investigate how the derivative interactions
affect cosmological solutions in this case, but 
it is beyond the scope of this paper.

\section{Conclusion}
In this paper, we discussed the possibility of adding 
consistent derivative interactions
in de Rham-Gabadadze-Tolley massive gravity.
On the construction of derivative interactions,
we required the following restrictions : 
(i) Linearization of $h_{\mu\nu}$ reproduces Fierz-Pauli massive gravity, 
(ii) the cutoff energy scale is $\Lambda_3$,
(iii) a derivative interaction term should contribute 
   at the energy scale $\Lambda_3$,
and (iv) the resultant theory does not have a Boulware-Deser ghost.
We found the derivative interactions, which 
reproduce the pseudo-linear derivative interaction term 
discussed in \cite{DI}.
These Lagrangians yield the problematic terms 
in the decoupling limit, whose energy scales are
below the cutoff energy scale $\Lambda_3$ in 
dRGT massive gravity.
We showed that these terms can be eliminated 
by adding the appropriate counterterms,
and the remaining terms in the decoupling limit
have the energy scale $\Lambda_3$.
Furthermore, 
we showed that 
the infinite number of derivative interactions can be resummed by using 
the tensor ${\cal K}^{\mu}_{~\nu}$,
which is the same tensor introduced in
de Rham-Gabadadze-Tolley massive gravity.
Thus, the resummed Lagrangian contains only two parameters.
However, the equations of motion for the scalar and tensor modes
in the decoupling limit are fourth order differential equations,
which generically implies the existence of Boulware-Deser ghosts.
Thus, the derivative interactions in de Rham-Gabadadze-Tolley
massive gravity, constructed under the above restrictions,
suffers from ghosts, which appear at $\Lambda_3$ 
in the decoupling limit. 
This implies that the mass scale of 
the derivative interactions has to be smaller than the Planck mass,
then Boulware-Deser ghosts of the derivative interactions 
can be pushed to a higher energy scale than $\Lambda_3$.

\vspace{10mm}

{\bf Acknowledgments:}
We would like to thank G. Gabadadze and S. Mukohyama for useful comments.
R.K. and D.Y. were supported in part by a Grant-in-Aid for JSPS Fellows.

\vspace{10mm}

\appendix

\section{Hamiltonian analysis}
In this appendix, for completeness, 
we use the Arnowitt-Deser-Misner formalism to show the existence of ghosts.

First we rewrite the Lagrangian in terms of 
$\Sigma^{\mu}_{~\nu}\equiv (\sqrt{g^{-1}\eta})^{\mu}_{~\nu}$ instead of ${\cal K}$.
Then the Lagrangian of the derivative interaction is given by
\begin{eqnarray}
  &&{\cal L}_{int}=-2(\alpha+\beta) M_{\rm Pl}^2\sqrt{-g} R\nonumber\\
  &&~~~~~~~~~
-M_{\rm Pl}^2\sqrt{-g} \varepsilon^{\mu\nu\rho\sigma}\varepsilon^{\alpha\beta\gamma\delta}
  R_{\mu\alpha\nu\beta}\, 
\biggl[
(\alpha+2\beta) g_{\rho\gamma}\Sigma_{\delta\sigma} -\beta \Sigma_{\rho\gamma}\Sigma_{\delta\sigma}
\biggr].
\end{eqnarray}
Here the first term is nothing but the Einstein-Hilbert term, 
which comes from derivative interactions.
Let us consider that the spacetime can be foliated by a family of spacelike hypersurfaces $\Sigma_t$,
defined by $t=x^0$\,.
The components of the metric can be parametrized by the lapse $N=1/\sqrt{-g^{00}}$, 
the shift $N^i=-g^{0i}/g^{00}$\,, and the metric on $\Sigma_t$ : $\gamma_{ij}=g_{ij}$
with $i,j$ running from $1$ to $3$.
We introduce the future-pointing unit normal vector $u^\mu$ to the surface $\Sigma_t$.
In terms of the ADM variables, the components of $u^\mu$ are
$u_0=-N,u_i=0,u^0=1/N,u^i=-N^i/N$\,.
The components of the spacetime metric are described as
\begin{eqnarray}
  &&g_{00} = -N^2 + N_i N^i, \qquad g_{0i}=N_i, \qquad g_{ij}=\gamma_{ij}\nonumber\\
  &&g^{00}=-{1 \over N^2}, \qquad g^{0i}={ N^i \over N^2}, 
  \qquad g^{ij}=\gamma^{ij}-{N^iN^j \over N^2},\nonumber
\end{eqnarray}
and $\sqrt{-g}=N\sqrt{\gamma}$. 
In the 3+1 decomposition, 
the projected Riemann tensor 
can be obtained by using the Gauss, Codazzi, and Ricci relations,
\begin{eqnarray}
&&R_{ijkl}
=K_{ik}K_{jl}-K_{il}K_{jk}+ {}^{(3)}R_{ijkl},\nonumber\\
&&R_{ijk\mu}u^\mu
=D_iK_{jk}-D_jK_{ik},\\
&&R_{i\mu j\nu}u^\mu u^\nu
=-N^{-1}\left(\partial_t K_{ij} - \pounds_{N^k} K_{ij}\right) + K_i{}^kK_{kj} + N^{-1}D_iD_j N,\nonumber
\end{eqnarray}
where $D_i$ denotes the covariant derivative associated with $\gamma_{ij}$\,,
${}^{(3)}R_{ijkl}$ is the Riemann tensor of $\gamma_{ij}$\,,
$K_{ij}$ is the extrinsic curvature, 
\begin{eqnarray}
  K_{ij}
= {1 \over 2N}\left(\partial_t\gamma_{ij}-D_iN_j-D_jN_i\right) ,
\end{eqnarray}
and 
$\pounds_{N^k} K_{ij} \equiv N^k D_k K_{ij} + K_{ik} D_j N^k + K_{jk}D_iN^k$. 
The square root tensor $\Sigma^\mu_{~\nu}$ can be decomposed as
\begin{equation}
\Sigma^\mu_{~\alpha}\Sigma^\alpha_{~\nu}={1\over N^2}
\begin{pmatrix}
1 & N^l\delta_{lj} \\
-N^i & (N^2 \gamma^{il}-N^iN^l)\delta_{lj}
\end{pmatrix}.
\end{equation}

In dRGT massive gravity, the Lagrangian is 
highly nonlinear in the lapse $N$,
which implies the lack of a Hamiltonian constraint.
However, one can introduce the new variable,
which is the combination of the lapse $N$ and shift $N^i$.
This new variable is independent of the lapse;
hence, the lapse $N$ becomes a Lagrange multiplier,
and a Hamiltonian constraint and 
an associated secondary constraint always exist.
In the extension theory of dRGT massive gravity,
we also require the existence of a Hamiltonian constraint
and a secondary constraint for avoiding a BD ghost. 
Since the combination of the lapse $N$ and shift $N^i$ should be 
determined by the mass term in dRGT theory,
we investigate the existence of 
a Hamiltonian constraint in the derivative interaction term
using the Hassan and Rosen method.

Following the proof by Hassan and Rosen \cite{HR},
we introduce a redefined lapse $n^i$,
\begin{eqnarray}
  N^i&=&(\delta^i_{~j}+N \,D^i_{~j})n^j.
\end{eqnarray}
Then the square root matrix can be written in the form
\begin{eqnarray}
  N \, \Sigma^\mu_{~\nu}&=& { A}^\mu_{~\nu} + N\,  {B}^\mu_{~\nu},
\end{eqnarray}
where $A^\mu_{~\nu}$ and $B^\mu_{~\nu}$ are given by
\begin{eqnarray}
  { A}^\mu_{~\nu}&=&{1\over \sqrt{x}}
  \begin{pmatrix}
    1 & n^l\delta_{lj} \\
    -n^i & -n^in^l\delta_{lj}
  \end{pmatrix},\\
  { B}^\mu_{~\nu}&=&
  \begin{pmatrix}
    0 & 0 \\
    0 & \sqrt{(\gamma^{il}-D^i_{~k}n^kD^l_{~m}n^m)\delta_{lj}}
  \end{pmatrix},\\
 \sqrt{x}D^i_{~j}&=&\sqrt{(\gamma^{il}-D^i_{~k}n^kD^l_{~m}n^m)\delta_{lj}},
 \label{DmatrixEQ}
\end{eqnarray}
where we defined $x \equiv 1-\delta_{ab}n^an^b$.
With this lapse $n^i$, 
the dRGT mass term is linear in the lapse $N$, 
which gives the Hamiltonian constraint.
However, we now have the time derivative of the extrinsic curvature
in the Riemann tensor. This might give a time derivative of
the lapse $N$ and shift $n^i$.
Therefore the first task is  to check whether lapse
and shift contain a time derivative or not.
A problematic term comes from $R_{i\mu j\nu}u^\mu u^\nu$ :
\begin{eqnarray}
  {\cal L}_{\rm int} \supset
-2M_{\rm Pl}^2N\sqrt{\gamma} 
\varepsilon^{iab}\varepsilon^{jcd}R_{i\mu j\nu}u^\mu u^\nu
\biggl[
(\alpha+2\beta) \gamma_{ac}\Sigma_{bd} -\beta \Sigma_{ac}\Sigma_{bd}
\biggr],
\end{eqnarray}
where $\varepsilon^{ijk}\equiv\varepsilon^{\mu ijk}u_\mu$. 
Then the $\partial_t K_{ij}$ term produces the time derivative of 
the lapse and shift.
This term is given by
\begin{eqnarray}
  {\cal L}_{\rm int} \supset
2  M_{\rm Pl}^2\sqrt{\gamma} 
\partial_t K_{ij}  \gamma^{ja}
\varepsilon^{ikl}\varepsilon_{abc}
\biggl[
(\alpha+2\beta) \delta^b_{~k}\Sigma^c_{~l} 
-\beta \Sigma^b_{~k}\Sigma^c_{~l}
\biggr].
\end{eqnarray}
In order to avoid the time derivative of the lapse $N$ and shift $n^i$,
we require that 
\begin{eqnarray}
  \varepsilon^{ikl}\varepsilon_{abc}
\biggl[
(\alpha+2\beta) \delta^b_{~k}\Sigma^c_{~l} 
-\beta \Sigma^b_{~k}\Sigma^c_{~l}
\biggr]={1 \over N^2} X^i_{~a} + {1 \over N} Y^i_{~a} + Z^i_{~a}
\end{eqnarray}
is independent of lapse $N$ and shift $n^i$.
First the $X^i_{~a}$ term is automatically zero ;
\begin{eqnarray}
X^i_{~a}=(\alpha+2\beta)
\biggl(
\delta^i_{~a} [A]^2 -\delta^i_{~a} [A^2]+ 2 A^i_{~b}A^b_{~a} -2 A^i_{~a} [A] 
\biggr)
=0.
\end{eqnarray}
In the second equality, 
we have used the fact that $A^i_j=[A] \, {\hat n}^i{\hat n}^l\delta_{lj}$,
where $[A] \equiv {\rm Tr} A^i{}_j=-(1-x)/\sqrt{x}$,
${\hat n}^i=n^i/\sqrt{1-x}$, and $\delta_{ab}{\hat n}^a{\hat n}^b=1$.
Before evaluating $Y^i_{~a}$, 
we decompose $B^i_j$ as 
\begin{eqnarray}
  B^i_{~j}={\hat B_0}{\hat n}^i{\hat n}^k\delta_{kj}+{\hat B}^i{\hat n}^k\delta_{kj}
+{\hat B}_{j}{\hat n}^i+{\hat B}^i_{~j}
\end{eqnarray}
where ${\hat B}_j {\hat n}^j=0$, ${\hat B}^i {\hat n}^k \delta_{ki}=0$, 
${\hat B}^i_{~j} {\hat n^j}=0$, and ${\hat B}^i_{~j} {\hat n}^k \delta_{ki}=0$.
Then the $Y^i_{~a}$ term becomes
\begin{eqnarray}
  Y^i_{~a}  = [A] \biggl[
\biggl\{(\alpha+2\beta)-2\beta [{\hat B}]\biggr\} P^i_{~a}
+2\beta {\hat B}^i_{~a}
\biggr]
\end{eqnarray}
where $[{\hat B}]\equiv{\rm Tr}{\hat B}^i{}_j$ and $P^i_{~j}$ is the projection tensor, 
$P^i_{~j} \equiv \delta^i_{~j}-{\hat n}^a{\hat n}^k\delta_{kj}$.
To eliminate the $Y^i_{~a}$ term, 
the only option is ${\hat B}^i_{~a}=b \,P^i_{~a}$, 
where $b$ is a constant; 
then,  
\begin{eqnarray}
  Y^i_{~a}  = [A]  P^i_{~a}
(\alpha+2\beta-2\beta b)
\end{eqnarray}
Thus the $1/N$ term cancels if $b=1+\alpha/2\beta$.
Now we have to check whether this solution satisfies
Eq.(\ref{DmatrixEQ}). 
To see this, we project the square of Eq.(\ref{DmatrixEQ});
then, we have three equations,
\begin{eqnarray}
  {1 \over x}{\hat B}_0^2 + {\hat B}_k{\hat B}^k
  &=&{\hat n}^a\delta_{ab}\gamma^{bc}\delta_{cd} {\hat n}^d\\
  {1 \over x}{\hat B}_0{\hat B}^i + {\hat B}^k{\hat B}^{i}_{~k}
  &=&{\hat n}^a\delta_{ab}\gamma^{bc}P^i_{~c}\\
  {1 \over x}{\hat B}^i{\hat B}_j + {\hat B}^i_{~k}{\hat B}^k_{~j}
  &=&\gamma^{ab}\delta_{bc}P^i_{~a}P^c_{~j}
\end{eqnarray}
Here we used the fact that $\delta_{ik}{\hat B}^k={\hat B}_i$, which 
can be obtained from $\delta_{ik}D^k_{~j}=\delta_{jk}D^k_{~i}$.
It is obvious that the solution of ${\hat B}^i_{~j}$ contains $\gamma_{ij}$, 
which means ${\hat B}^i_{~j} \neq P^i_{~j}$.
This can be checked by substituting the ansatz, 
${\hat B}^i_{~a}=b \,P^i_{~a}$.
Thus $Y^i_{~j}$ is not zero, and the lapse $N$ and the shift $n^i$
are dynamical variables. 

However, it should be stressed that 
the new shift variable is introduced so that
the dRGT mass term yields a Hamiltonian constraint.
Therefore, there might exist a way to avoid dynamical lapse 
or shift by introducing new variables, but this is still an unsolved problem.
Therefore, this proof only shows the lack of 
Hamiltonian and momentum constraints.

\vspace{10mm}


\end{document}